\documentclass[twocolumn,preprintnumbers,amsmath,amssymb]{revtex4}
\usepackage{docs}

\usepackage{graphicx}
\usepackage{dcolumn}

\usepackage{bm}

\def\frac#1#2{{\textstyle{{#1}\over {#2}}}}
\def\goto{\rightarrow}

\def\lsim{\mathrel{\rlap{\lower4pt\hbox{\hskip1pt$\sim$}}
    \raise1pt\hbox{$<$}}}
\def\gsim{\mathrel{\rlap{\lower4pt\hbox{\hskip1pt$\sim$}}
    \raise1pt\hbox{$>$}}}
\def\sqr#1#2{{\vcenter{\vbox{\hrule height.#2pt
         \hbox{\vrule width.#2pt height#1pt \kern#1pt
         \vrule width.#2pt}
         \hrule height.#2pt}}}}

\def\beq{\begin{equation}}
\def\eeq{\end{equation}}
\def\beqa{\begin{eqnarray}}
\def\eeqa{\end{eqnarray}}

\def\laq{\raise 0.4 ex \hbox{$<$}\kern -0.8 em\lower 0.62 ex\hbox{$\sim$}}
\def\gaq{\raise 0.4 ex \hbox{$>$}\kern -0.7 em\lower 0.62 ex\hbox{$\sim$}}

\begin{document}
%-----------------------------------------------
\title{Time and Causation}%
%-----------------------------------------------

\author{Orfeu Bertolami}
\email{orfeu@cosmos.ist.utl.pt} \affiliation{Instituto Superior
T\'ecnico, Departamento de F\'\i sica, \\ Instituto de Plasmas e
Fus\~ao Nuclear, \\ Av. Rovisco Pais 1, 1049-001 Lisboa, Portugal}

\author{Francisco S. N. Lobo}
\email{flobo@cosmo.fis.fc.ul.pt}
%\affiliation{Institute of
%Cosmology \& Gravitation,
%             University of Portsmouth, Portsmouth PO1 2EG, UK}
\affiliation{Centro de Astronomia e Astrof\'{\i}sica da
Universidade de Lisboa, Campo Grande, Ed. C8 1749-016 Lisboa,
Portugal}

\date{\today}

%-----------------------------------------------
\begin{abstract}
%-----------------------------------------------

In this review paper, we consider the fundamental nature of time
and causality, most particularly, in the context of the theories
of special and general relativity. We also discuss the issue of
closed timelike curves in the context of general relativity, and
the associated paradoxes, the question of directionality of the
time flow and, rather briefly, the problem of time in quantum
gravity.

%-----------------------------------------------
\end{abstract}
%-----------------------------------------------

%\pacs{04.50.+h, 04.20.Gz}

%-----------------------------------------------
\maketitle
%-----------------------------------------------

\noindent
{\it Time present and time past}

\noindent
{\it Are both perhaps present in time future}

\noindent
{\it And time future contained in time past.}

\noindent
{\it If all time is eternally present}

\noindent
{\it All time is unredeemable.}

\noindent 
{\it What might have been is an abstraction}

\noindent
{\it Remaining perpetual possibility}

\noindent 
{\it Only in a world of speculation.}

\noindent 
{\it What might have been and what has been}

\noindent 
{\it Point to one end, which is always present.}

\vspace{0.2cm}

\noindent
{ Burt Norton}

\noindent {T.S. Eliot}

%-----------------------------------------------
\section{Introduction}
%-----------------------------------------------

Time is a most mysterious ingredient of the Universe and does
stubbornly resist any simple definition. Intuitively, the notion
of time seems to be intimately related to change, and subjectively
it is clearly perceived as something that flows. This view can be
traced back as far as Aristotle (384 BC - 322 BC), a keen natural
philosopher, who categorically stated that {\it time is the measure
of change}. Throughout history, one may find a wide variety of
reflections and considerations on time, dating back to ancient
religions. For many civilizations, cycles in nature were an
evidence of the circular nature of time. Indeed, it was only in
the 17th century that the philosopher Francis Bacon (1561 - 1626)
clearly formulated the concept of linear time, and through the
influence of Newton (1643 - 1727), Barrow (1630 - 1677), Leibniz
(1646 - 1716), Locke (1632 - 1704) and Kant (1724 - 1804) amongst
others, by the 19th century the idea of a linear time evolution
was a dominant one both in science and philosophy (a very partial
list of references include
\cite{Russell,Prigogine1,Montheron,Ellis,Tipler80,Coveney,Lobo:2007ms,Bertolami:2008ps,Bertolami06b}).

In a scientific context, it is perhaps fair to state that
reflections on time culminated with Newton's concept of absolute
time, which assumed that time flowed at the same rate for all
observers in the Universe \cite{Newton}. Newton compared time and
space with an infinitely large vessel, containing events, and
existing independently of the latter. However, in 1905, Albert
Einstein (1879 - 1955) changed altogether our notion of time,
through the formulation of the special theory of relativity and
stating, in particular, that time flowed at different rates for
different observers. Three years later, Hermann Minkowski (1879 -
1909) \cite{Minkowski} suggested the unification of the time and
space parameters, giving rise to the notion of a fundamental
four-dimensional entity, {\it spacetime} (see e.g. \cite{Petkov}
for an extensive discussion). Furthermore, in 1915, Einstein put
forward the general theory of relativity where it was shown that
the {\it flat} or {\it uncurved} spacetime of special relativity
is curved by energy/matter. Since then the discovery of new forms
of energy/matter in physics has often given rise to new spacetime
geometries. Physics and geometry are once again intertwined 
(see for instance \cite{Bertolami06a}).

In this context, an interesting description of spacetime is the
so-called {\it Block Universe} which represents spacetime as an
unchanging four-dimensional block, where time is considered a
dimension. In this representation, a preferred {\it present} is
non-existent and past and future times are equally present. All
points in time are equally valid frames of reference, and whether
a specific instant is in the future or past is frame dependent.
However, despite the fact that each observer does indeed
experience a subjective flow of time, special relativity denies
the possibility of universal simultaneity, and therefore the
possibility of an universal present. We refer the reader to Ref.
\cite{Ellis:2006sq} for details on the objections to the Block
Universe viewpoint.

An important aspect of the nature of time concerns its flow. The
modern physical perspective regards the Universe as described by
{\it dynamical laws}, from which, after specifying a suitable set of
initial conditions for a physical state, the time evolution of the
system is determined. The fundamental dynamical equations of
classical and quantum physics are symmetrical under a time
reversal, i.e., mathematically, one might as well specify the
final conditions and evolve the physical system back in time.
However, in macroscopic phenomena, which are accurately
described by thermodynamics, as well as some instances in general
relativity and quantum mechanics, the evolution of the systems
seems to be essentially time asymmetric. This enables observers to
empirically distinguish past from future. Indeed, the Second Law
of Thermodynamics, which states that in an isolated system the
entropy (which is a measure of disorder) increases provides a
direction for the flow of time. It is an interesting possibility
that the Second Law of Thermodynamics and the thermodynamic arrow
of time are a consequence of the initial conditions of the
Universe, which sets a cosmological direction for the flow of
time, that inexorably points in the evolution flow of the
Universe's expansion.

One should notice that as time is incorporated into the very
fabric of spacetime, concern should arise from the fact that
general relativity is contaminated with non-trivial geometries
that generate closed timelike curves, and apparently violates
causality. A closed timelike curve allows time travel, in the
sense that an observer who travels in spacetime along this curve,
returns to an event that coincides with the departure. This fact
apparently violates causality and produces time travel paradoxes
\cite{Nahin}. The notion of causality is fundamental in the
construction of physical theories; therefore time travel and its
associated paradoxes have to be treated with great caution
\cite{Visser}.

This review paper is outlined in the following way: In Section
\ref{Sec:II}, we consider the fundamental nature of time in
special and general relativity, paying close attention to the time
dilation effects. In Section \ref{Sec:III}, we address the issue
of closed timelike curves and the violation of causality, and in
Section \ref{Sec:IV} we discuss the issue of directionality of the
time flow. In Section \ref{Sec:V}, we present some open issues,
such as the correlation of the arrows of time, and the problem of
time in quantum gravity. Finally in Section \ref{Sec:Conclusion},
we present our conclusions.

\section{Relativistic time}\label{Sec:II}

The conceptual definition and understanding of time, both
quantitatively and qualitatively is somewhat complex. Special
relativity provides us the framework to quantitatively address the
fundamental processes related to time dilation effects. The
general theory of relativity, on its hand, accounts for the
effects on the flow of time in the presence of gravitational
fields. Both, special and general relativity are most successful
theories from the experimental point of view. General relativity,
for instance, is well established in the weak gravitational field
limit \cite{Will,BPT}. Its predictions range from the existence of
black holes, gravitational radiation to the cosmological models
predicting a primordial beginning, the Big-Bang
\cite{HawkingEllis,Wald}.

\subsection{Time in special relativity}

In 1905, Einstein abandoned the postulate of absolute time, and
assumed instead the following two postulates: $(i)$ the speed of
light, $c$, is the same in all inertial frames; $(ii)$ the
{\it principle of relativity, which states that the laws of physics
take the same form in every inertial frame}. Considering an
inertial reference frame ${\cal O}'$, with coordinates
$(t',x',y',z')$, moving along the $x$ direction with uniform
velocity, $v$, relative to another inertial frame ${\cal O}$, with
coordinates $(t,x,y,z)$, and taking into account the above two
postulates, Einstein deduced the Lorentz transformation, the
transformations relating the two coordinate systems, which are
given by
\begin{eqnarray}\label{Lorentz}
t'&=&\gamma (t-vx/c^2) \,,\nonumber  \\
x'&=&\gamma (x-vt) \,, \\
y'&=&y\,,\qquad  z'=z\,,\nonumber
\end{eqnarray}
where $\gamma$ is defined as $\gamma=(1-v^2/c^2)^{-1/2}$. One
immediately verifies, from the first two equations, that the time
and space coordinates are mixed by the Lorentz transformation, and
hence, the viewpoint that the physical world is modelled by a
four-dimensional spacetime continuum.

Considering now two events, $A$ and $B$, respectively, with
coordinates $(t_A,x_A,y_A,z_A)$ and $(t_B,x_B,y_B,z_B)$ in an
inertial frame ${\cal O}$, then the interval between the events is
given by
\begin{equation}
\Delta s^2=-c^2\Delta t^2+\Delta x^2+\Delta y^2+\Delta z^2 \,,
\label{SpecRel}
\end{equation}
where $\Delta t$ is the time
interval between the two events $A$ and $B$ \cite{Hobson}. One
verifies that the expression (\ref{SpecRel}) is invariant under
Lorentz transformations, and as advocated by Minkowski, space and
time are united in a four-dimensional entity, {\it spacetime}.
Thus, the interval (\ref{SpecRel}) may be considered as an
underlying geometrical property of the spacetime itself, actually
the distance between points (events) in spacetime.

The sign $\Delta s^2$ is also invariantly defined, so that $\Delta
s^2<0$ is a timelike interval; $\Delta s^2=0$, a null interval;
and $\Delta s^2>0$, a spacelike interval. Observers moving with a
relative velocity $v<c$ travel along timelike curves, which are
referred to as the {\it worldline} of the observer. There is now a
unique time measured along a worldline, the {\it proper time}. A
photon travels along null curves.

The special theory of relativity challenges many of our intuitive
beliefs about time. For instance, the theory is inconsistent with
the common belief that the temporal order in which two events
occur is independent of the observer's reference frame. Thus,
whether a specific instant is in the future or past is frame
dependent. {\it Special relativity rules out the possibility of
universal simultaneity, and hence the possibility of a universal
present}.

Another of our intuitive beliefs challenged by the special theory
of relativity is related with the time dilation effects. Let us
exemplify this issue. Suppose that a clock sits at rest with
respect to the inertial reference frame ${\cal O}'$, in which two
successive clicks, represented by two events $A$ and $B$ are
separated by a time interval $\Delta t'$. To determine the time
interval $\Delta t$ as measured by ${\cal O}$, it is useful to
consider the inverse Lorentz transformation, given by
\begin{equation}
t=\gamma (t'+vx'/c^2)\,,
\end{equation}
which provides
\begin{equation}
t_B-t_A=\gamma \left[t'_B-t'_A+v(x'_B-x'_A)/c^2\right]\,,
\end{equation}
where $t_A$ and $t_B$ are the two clicks measured in ${\cal O}$.
As the events are stationary relative to ${\cal O}'$, we have
$x'_B=x'_A$, so that one finally ends up with
\begin{equation}
\Delta t=\gamma \Delta t'\,,
\label{timedilation}
\end{equation}
where $\Delta t=t_B-t_A$ and $\Delta t'=t_B'-t_A'$. As $\gamma
>1$, then $\Delta t>\Delta t'$, so that time as measured by the
moving reference frame ${\cal O}'$ slows down relatively to ${\cal
O}$. This feature has been observed experimentally, in particular,
in the Hafele-Keating experiment performed on October 1971
\cite{Hafele-Keating}. We note that the fact that a moving clock
slows down is completely reciprocal for any pair of inertial
observers, and this is essentially explained as both disagree
about simultaneity.

\subsection{Time in general relativity}

The analysis outlined above has only accounted for flat
spacetimes, contrary to Einstein's general theory of relativity,
in which gravitational fields are accounted through the curvature
of spacetime. In the discussion of special relativity, the
analysis was restricted to inertial motion, but in general
relativity the principle of relativity is extended to all
observers, inertial or non-inertial. In general relativity it is
assumed that the {\it the laws of physics are the same for all
observers, irrespective of their state of motion}. However, given
that a gravitational force measured by an observer depends on his
state of acceleration, one is led to the {\it principle of
equivalence}, which states that {\it there is no way of
distinguishing the effects on an observer of a uniform
gravitational field from the ones of a constant acceleration}.

As already mentioned, general relativity is a quite well
established theory from the experimental view point. For instance,
the global positioning system (GPS) would not work at all without
the general relativistic corrections to Newtonian mechanics and
gravity (see for instance, \cite{Paramos08} and references therein).
Actually, likewise in special relativity, one expects
time dilation effects now due to gravitational fields. In order to
exemplify this imagine the following idealized thought experiment,
actually suggested by Einstein himself \cite{Schutz}. Consider a
tower of height $h$ hovering on the Earth's surface, with a
particle of rest mass $m$ lying on top. The particle is then
dropped from rest, falling freely with acceleration $g$ and
reaches the ground with a non-relativistic velocity
$v=(2gh)^{1/2}$. Thus, an observer on the ground measures its
energy as
\begin{equation}\label{energy}
E=mc^2+\frac{1}{2}mv^2=mc^2+mgh \,.
\end{equation}
The idealized particle is then converted into a single photon
$\gamma_1$ with identical energy $E$, which returns to the top of
the tower. Upon arrival it converts into a particle with energy
$E'=m'c^2$. Notice that in order to avoid perpetual motion, $m'>m$
is forbidden, hence, we consider $m=m'$, and the following
relationship is obtained
\begin{equation}\label{E-ratio}
{E' \over E}={mc^2 \over mc^2+mgh}\simeq 1-{gh \over c^2} \,,
\end{equation}
as $gh/c^2\ll 1$. From the definitions $E=h\nu$ and $E'=h\nu'$,
where $\nu$ and $\nu'$ are the frequencies of the photon at the
bottom and top of the tower, then from Eq. (\ref{E-ratio}), one
obtains
\begin{equation}\label{nu-ratio}
\nu'=\nu \left(1-{gh \over c^2}\right) \,.
\end{equation}
Now, in order to get the result that clocks run at different rates in a
gravitational field, consider the following {\it gedanken} or
{\it thought experiment}.
The observer at the bottom of the tower emits a light wave,
directed to the top. The relationship of time between two crests
is simply the inverse of the frequency, i.e., $\Delta t=1/\nu$, so
that from Eq. (\ref{nu-ratio}), one obtains in the approximation $gh/c^2\ll 1$:
\begin{equation}\label{GRtimedilate}
\Delta t'=\Delta t \left(1+{gh \over c^2}\right) \,.
\end{equation}
This clearly shows that time flows at a faster rate on top
of the tower than at the bottom. Note that this result has been
obtained independently of the gravitational theory.

Actually, this experiment is a well-known test of the time
dilation effects in general relativity, first preformed by Pound
and Rebka \cite{Pound-Rebka}, which confirmed the predictions of
general relativity to a $10\%$ precision level
\cite{Pound-Rebka2}. These results were subsequently improved
to a $1\%$ precision level by Pound and Snider
\cite{Pound-Snider}. To within experimental errors, all
experimental results are consistent with the special and general
relativistic predictions.

It is remarkable that any particle, however elementary, is subjected to 
gravity as described above even at quantum level, as 
recently proved experimentally 
for ultra-cold neutrons \cite{Nesvizhevsky}. This experiment is also 
consistent with the equivalence principle \cite{BNunes03}.

\section{Closed timelike curves and causality violation}\label{Sec:III}

As time is incorporated into the very structural fabric of
spacetime, it is interesting to note that general relativity is
contaminated with non-trivial geometries which generate {\it
closed timelike curves} \cite{Visser}. A closed timelike curve
(CTC) allows time travel, in the sense that an observer which
travels on a trajectory in spacetime along this curve, returns to
an event which coincides with the departure. The arrow of time
leads forward, as measured locally by the observer, but globally
he/she may return to an event in the past. This fact apparently
violates causality and produces time travel paradoxes
\cite{Nahin}. The notion of causality is fundamental in the
construction of physical theories, therefore time travel and its
associated paradoxes have to be treated with great care. These
paradoxes fall into two broad groups, namely the {\it consistency
paradoxes} and the {\it causal loops}.

The consistency paradoxes include the classical grandfather
paradox. Imagine traveling into the past and meeting one's
grandfather. Nurturing homicidal tendencies, the time traveler
murders his/her grandfather, impeding the birth of his/her father,
therefore making his/her own birth impossible. The consistency
paradoxes occur whenever possibilities of changing events in the
past arise.

The paradoxes associated with causal loops are related to
self-existing information or objects, trapped in spacetime.
Imagine a researcher getting the full formulation of a consistent
theory of quantum gravity from a time traveler from the future.
He/she eventually publishes the article in a high-impact journal
and as the years advance, he/she eventually travels to the past
providing the details of the consistent quantum gravity theory
to a younger version of himself/herself. The article on the theory
of quantum gravity exists in the future because it was written in
the past by the young researcher. The latter wrote it up, after
receiving the details from his older version. Both parts
considered by themselves are consistent, and the paradox appears
when its elements are considered together. One is liable to ask,
what is the origin of the information, as it seems to arise out of
nowhere. The details for a complete and consistent theory of
quantum gravity, which paradoxically were never created,
nevertheless exist in spacetime. Note the absence of causality
violations in these paradoxes.

A great variety of solutions to the Einstein field equations
containing CTCs exist, but two particularly
notorious features seem to stand out \cite{Lobo:2002rp}. Solutions
with a tipping over of the light cones due to a rotation about a
cylindrically symmetric axis \cite{rotating}; and solutions that
violate the energy conditions of general relativity, which are
fundamental in the singularity theorems and theorems of classical
black hole thermodynamics \cite{Visser,HawkingEllis}.

\subsection{Solutions violating the energy conditions}

The usual way to obtain solutions of the Einstein field equations
consists in considering a plausible distribution of energy/matter,
and then find the resulting geometrical structure. However, one
can run the Einstein field equation in the reverse direction by
imposing an exotic geometrical spacetime structure, and eventually
determining the energy/matter source for that geometry.

In this fashion, solutions violating the energy conditions have
been obtained. One of the simplest energy conditions is the {\it
weak energy condition}, which is essentially equivalent to the
assumption that any timelike observer measures a positive local
energy density. Although classical forms of matter obey these
energy conditions, violations have been encountered in quantum
field theory, the Casimir effect being a well-known example. By
adopting the reverse procedure, solutions such as traversable
wormholes \cite{Morris,Visser,WHsolutions}, the warp drive
\cite{Alcubierre,Lobo:2004wq,Lobo:2002zf}, and the Krasnikov tube
\cite{Krasnikov} have been obtained. These solutions violate the
energy conditions and through rather simple manipulations generate
CTCs \cite{MT,Everett,ER,frolovnovikovTM}.

We briefly consider here the specific case of {\it traversable
wormholes} \cite{Morris}. A wormhole is essentially constituted by
two mouths, $A$ and $B$, residing in different regions of
spacetime \cite{Morris}, which in turn are connected by a tunnel
or handle. One of the most fascinating aspects of wormholes is
actually how easily they allow for generating CTCs \cite{MT}.
There are several ways to generate a time machine using multiple
wormholes \cite{Visser}, but the manipulation of a single wormhole
is the simplest way \cite{MT}. The basic idea is to create a time
shift between both mouths. This is done invoking the time dilation
effects of special or general relativity, i.e., one may consider
the analogue of the twin paradox, in which the mouths are moving
one with respect to the other, or instead, the case in which one
of the mouths is placed in a strong gravitational field.

To create a time shift using the twin paradox analogue, consider
that the mouths of the wormhole may be moving one with respect to
the other in external space, without significant changes of the
internal geometry of the handle. For simplicity, consider that one
of the mouths $A$ is at rest in an inertial frame, whilst the
other mouth $B$, initially at rest when close by to $A$, but starts
moving out with a high velocity and then returns to its starting point.
Due to the Lorentz time contraction, the time interval between
these two events, $\Delta T_B$, measured by a clock comoving with
$B$ can be made to be significantly shorter than the time interval
between the same two events, $\Delta T_A$, as measured by a clock
resting at $A$. Thus, the clock that has moved has been slowed by
$\Delta T_A-\Delta T_B$ relative to the standard inertial clock.
Note that the tunnel, between $A$ and $B$ remains
practically unchanged, so that an observer comparing the time of
the clocks through the handle will measure an identical time, as
the mouths are at rest with respect to one another. However, by
comparing the time of the clocks in external space, he/she will
verify that the time shift is precisely $\Delta T_A-\Delta T_B$,
as both mouths are in different reference frames. Now, consider an
observer starting off from $A$ at an instant $T_0$, measured by
the clock at rest in $A$. He/she makes his/her way to $B$ in
external space and enters the tunnel from $B$. Consider, for
simplicity, that the trip through the wormhole tunnel is
instantaneous. He/she then exits from the wormhole mouth $A$ into
external space at the instant $T_0-(\Delta T_A-\Delta T_B)$ as
measured by a clock positioned at $A$. His/her arrival at $A$
precedes his/her departure, and the wormhole has been converted
into a time machine.

For concreteness, following the analysis of Morris {\it et al}
\cite{MT}, we consider the metric of the accelerating
wormhole given by
\begin{eqnarray}
ds^2&=&-(1+glF(l)\cos\theta)^2\;e^{2\Phi(l)}\;dt^2+dl^2
     \nonumber  \\
&&+r^2(l)\,(d\theta^2 +\sin^2\theta\,d\phi^2)   \,,
    \label{accerelatedWH}
\end{eqnarray}
where the proper radial distance, $dl=(1-b/r)^{-1/2}\,dr$, is
used. $F(l)$ is a form function that vanishes at the wormhole
mouth $A$, at $l\leq 0$, rising smoothly from 0 to 1, as one moves
to mouth $B$; $g=g(t)$ is the acceleration of mouth $B$ as
measured in its own asymptotic rest frame. Consider that the
external metric to the respective wormhole mouths is $ds^2 \cong
-dT^2+dX^2+dY^2+dZ^2$. Thus, the transformation from the wormhole
mouth coordinates to the external Lorentz coordinates is given by
\begin{eqnarray}
T&=&t\,, \qquad Z=Z_A+l\,\cos\theta\,,
   \nonumber  \\
X&=&l\,\sin\theta\,\cos\phi \,, \qquad  Y=l\,\sin\theta\,\sin\phi
\,,
\end{eqnarray}
for mouth $A$, where $Z_A$ is the time-independent $Z$ location of
the wormhole mouth $A$, and
\begin{eqnarray}
T&=&T_B+v\gamma \,l\,\cos\theta\,, \qquad
Z=Z_B+\gamma\,l\,\cos\theta\,,
     \nonumber   \\
X&=&l\,\sin\theta\,\cos\phi \,, \qquad Y=l\,\sin\theta\,\sin\phi
\,,
\end{eqnarray}
for the accelerating wormhole mouth $B$. The world line of the
center of mouth $B$ is given by $Z=Z_B(t)$ and $T=T_B(t)$ with
$ds^2=dT_B^2-dZ_B^2$; $v(t)\equiv dZ_B/dT_B$ is the velocity of
mouth $B$ and $\gamma=(1-v^2)^{-1/2}$ the respective Lorentz
factor; the acceleration appearing in the wormhole metric is given
$g(t)=\gamma^2\;dv/dt$~\cite{MT}.

Novikov considered other variants of inducing a time shift through
the time dilation effects in special relativity, by using a
modified form of the metric (\ref{accerelatedWH}), and by
considering a circular motion of one of the mouths with respect to
the other~\cite{Novikov-CTCWH}. Another interesting way to
induce a time shift between both mouths is simply to place one of
the mouths in a strong external gravitational field, so that times
slows down in the respective mouth. The time shift will be given
by $T=\int_{i}^{f}\,(\sqrt{g_{tt}(x_A)}-\sqrt{g_{tt}(x_A)}\;)\;dt$
~\cite{Visser,frolovnovikovTM}.

\subsection{Possible solutions of the time travel paradoxes?}

In what concerns the solution of the violation of causality, if
one regards that general relativity is a valid theory, then it is
plausible to at least include the {\it possibility} of time travel
in the form of CTCs. However, a caution reaction
is to exclude time travel due to the associated paradoxes,
although the latter do not prove that time travel is
mathematically or physically impossible. The paradoxes do indeed
indicate that local information in a spacetime containing CTCs is
restricted to rather unfamiliar situations. In what regards to the
grandfather paradox, it is logically inconsistent that the time
traveler murders his/her grandfather. But, one can ask, what
exactly prevents him/her from accomplishing the murderous act
given the opportunities and the free-will to do so. It seems that
certain conditions in local events have to be fulfilled, for the
solution to be globally self-consistent. These conditions are
referred to as {\it consistency constraints} \cite{Earman}. Much
has been written on two possible remedies to the CTC paradoxes,
namely the {\it Principle of Self-Consistency} and the {\it
Chronology Protection Conjecture}.

Novikov's {\it Principle of Self-Consistency} stipulates that
events on a CTC are self-consistent, i.e.,
events influence one another along the curve in a cyclic and
self-consistent way. In the presence of CTCs the
distinction between past and future events are ambiguous, and the
definitions considered in the causal structure of well-behaved
spacetimes break down. What is important to note is that events in
the future can influence, but cannot change, events in the past.
According to this principle, the only solutions of the laws of
physics that are locally allowed, and reinforced by the
consistency constraints, are those which are globally
self-consistent.

Hawking's {\it Chronology Protection Conjecture} is a more
conservative way of dealing with the causality paradoxes. Hawking
puts forward his conjecture based on the strong experimental
evidence that ``we have not been invaded by hordes of tourists
from the future'' \cite{Hawk2}. An analysis reveals that the value
of the renormalized expectation quantum stress-energy tensor
diverges close to the formation of CTCs, which
destroys the wormhole's internal structure before attaining the
Planck scale. There is no convincing demonstration of the {\it
Chronology Protection Conjecture}, but perhaps the expected
answers will arise from the quantum gravity theory.

\section{Arrows of time}\label{Sec:IV}

By the second half of the XIX century, the development of the
kinetic theory of matter by Maxwell (1831 - 1879), Clausius (1822
- 1888) and Boltzmann (1844 - 1906) did revive the discussion on
the problem of linear time evolution and of the eternal recurrence
of motion.

The idea of a cyclic time and of an eternal return was discussed
by the philosophers Herbert Spencer (1820 - 1903) and Friedrich
Nietzsche (1844 - 1900) about the same time that Poincar\'e (1854
- 1912) showed his fundamental theorem. Of course, their arguments
are not at the level of rigour as in physics and mathematics, but,
interestingly, the ``proof'' of Nietzsche contains elements which
are relevant for any discussion of the subject, such as a finite
number of states, finite energy, no creation of the universe and
chance-like evolution.

The starting point of physical discussion is the fact that
Newton's equations have no intrinsic time direction, being
invariant under time reversal. Nevertheless, Poincar\'e showed in
1890, in the context of classical mechanics, a general recurrence
theorem, according to which any isolated system, which includes
the universe itself, would return to its initial state given a
sufficiently long time interval.

Poincar\'e's theorem is valid in any space $X$ where there exists
a one parameter map $T_i$ from sets $[U]$ and a measure $\mu$ on
$X$ such that: i) $\mu(X)=1$ and ii)
$\mu(T_{t_0}(U))=\mu(T_{t_0+t}(U))$ for any subset of $X$ and any
$t_0$ and $t$. In classical mechanics, condition i) is ensured by
demanding that space $X$ is the phase space of a finite energy
system in a finite box. If $\mu$ is the distribution or density
function, $\rho$, in phase space and $T_t$ is the evolution
operator of the mechanical system (the Hamiltonian or the
Liouville operator), then condition ii) follows from Liouville's
theorem: $d\rho /dt=0$. Hence, it implies that classical mechanics
is inconsistent with the Second Principle of Thermodynamics.

Naturally, the recurrence theorem was a major issue in Boltzmann's
approach to the problem of irreversibility. Indeed, in the 1870s,
he realized that deducing an arrow of time from the mechanics of
atoms was impossible without using averaging arguments. The
developed formalism allowed him to understand statistical
equilibrium with the Liouville equation and in 1872 he obtained a
time-asymmetric evolution equation, now referred to as the
Boltzmann equation, whose solution was a single-particle
distribution function of a molecule in a diluted gas. From this
distribution function he could construct a strictly decreasing
function of time, the so-called ${\cal H}$-function. The
identification of the ${\cal H}$-function with minus the entropy,
allowed him to claim to have solved the irreversibility problem at
molecular level.

A crucial point however, was that in order to arrive at his result
Boltzmann had to rely on the assumption that molecules about to
collide are uncorrelated, but that after the collision they are
correlated as their trajectories are altered by the collision, the
{\it molecular chaos hypothesis} or {\it Stosszahlansatz}.
However, in 1876, Johann Josef Loschmidt (1821 - 1895), a friend
of Boltzmann, argued that the time-asymmetry obtained by Boltzmann
was entirely due to the time-asymmetry of the molecular chaos
assumption. Twenty years later, Ernest Zermelo (1871 - 1953), a
young assistant of Planck (1858 - 1947) in Berlin, attacked
Boltzmann again, now armed with Poincar\'e's recurrence theorem.
Boltzmann tried to save his case via a cosmological model. He
proposed that as a whole the universe had no time direction, but
that time-asymmetry could arise in some regions when through a
large fluctuation from equilibrium it would yield states of
reduced entropy. These regions of low entropy would evolve back to
the most likely state of maximum entropy, and the process would
then follow Poincar\'e's theorem. We know today that Poincar\'e's
theorem cannot be applied to the whole Universe given the
existence of spacetime singularities in general relativity.

By 1897, Planck started tackling the irreversibility problem in a
series of papers which actually culminated with his discovery of
the quantum theory of radiation in 1900. It was clear that a
finite system of particles would be recurrent and not irreversible
in the long run, and for this reason he considered instead a field
theory, electrodynamics. The hope was to derive irreversibility
from the interaction of a continuous field with a set of
particles. Planck's arguments led Boltzmann to remark that as a
field could be seen as a mechanical system with an infinite number
of molecules, an infinite Poincar\'e recurrence period should then
be expected, and thus a long term agreement with the observed
irreversibility from which the Second Principle would follow.

Despite of that the persistent objections of influential opponents
such as Ernest Mach (1838 - 1916) and Friedrich Ostwald (1853 -
1932), led Boltzmann into depression and into a first suicide
attempt in Lepzig, before assuming Mach's chair in Vienna in 1902.
The intellectual isolation and the continuous deterioration of his
health led him eventually to suicide at the age of 62, at Duino, a
seaside holiday resort on the Adriatico coast near Trieste, on the
5th September 1906.

Even after Boltzmann, the irreversibility problem has resisted any
simple explanation. In 1907, the couple Ehrenfest, Paul Ehrenfest
(1880 - 1933) and Tatyana Afanasyeva (1876 - 1964) further
developed Boltzmann's idea of averaging over a certain region,
$\Delta$, of the phase space and showed that the averaged ${\cal
H}$-function would remain strictly decreasing in the
thermodynamical limit, after which $\Delta$ could be taken as
small as compatible with the uncertainty principle (see e.g. Ref.
\cite{Huang}).

In 1928, Pauli (1869 - 1958) when considering the problem of
transitions in the context of quantum mechanical perturbation
theory showed that satisfying the Second Principle of
Thermodynamics would require a {\it master equation}:

\beq {d p_i \over d t} = \sum_j ~(\omega_{ij} p_j -\omega_{ji}
p_i) ~, \eeq where $\omega_{ij}$ is the conditional probability
per unit of time of the transition $j \goto i$ and $p_i$ is the
probability of state $i$. Assuming the {\cal H}-function to be
given by

\beq {\cal H} = \sum_i ~p_i \ln p_i ~, \eeq it follows that ${d
{\cal H} \over d t} \leq 0$. This approach is fairly suggestive as
it makes clear that that irreversible phenomena should be
understood in the context of microscopic physics, the very essence
of Boltzmann's work.

More recently, Prigogine (1917 - 2003) and collaborators put
forward a quite radical idea, according to which the
irreversible behaviour should be already incorporated at the
microscopic level (see e.g. Ref. \cite{Prigogine2} for a general
discussion). Mathematically, the problem consists in turning time
into an operator which does not commute with the Liouville
operator, itself the commutator of the Hamiltonian with the
density matrix. Physically, this proposal implies that reversible
trajectories cannot be considered, leading to an entropy-like
quantity which is a strictly increasing function of time.

One should realize that besides the difficulty in explaining from
microphysics the irreversible behaviour of macroscopic
systems, there exists in nature a variety of phenomena whose
behaviour exhibit an immutable flow from past to present, from
present to future. The term ``arrow of time'', coined by the
British astrophysicist and cosmologist Arthur Eddington (1882 -
1944) \cite{Eddington}, is often used to refer to this
evolutionary behaviour.  Let us briefly describe these phenomena
and their main features:

\vspace{0.3cm}

\noindent 1) The time asymmetry inferred from the growth of
entropy in irreversible and dissipative macroscopic phenomena as discussed
above.

\vspace{0.3cm}

\noindent 2) The nonexistence of advanced electromagnetic
radiation, coming from the infinite and converging to a source,
even though this a possible solution of the Maxwell's field
equations.

\vspace{0.3cm}

\noindent 3) The measurement process and the ensued collapse of
the wave function of a quantum system and the irreversible
emergence of the classical behaviour, despite the fact that the
fundamental equations of quantum mechanics and statistical quantum
mechanics, Schr\"odinger's and Von Neumann's equations,
respectively, are - likewise Newton's equation of motion -
invariant under time inversions for systems described by a
time-independent Hamiltonian (see e.g. Ref. \cite{Penrose1990,Penrose2005}
for a through discussion).

\vspace{0.3cm}

\noindent 4) The exponential degradation in time of systems and
the exponential growth of self-organized systems (for a
sufficiently abundant supply of resources). In the development of
self-organized systems, an important role is played by complexity.
The fascinating aspects of complex phenomena has led some authors to
refer to their rather unique evolution behaviour as ``creative
evolution'', ``arrow of life'' or ``physics of becoming''
\cite{Coveney,Prigogine1,Prigogine2}. In the context of complex
systems, the chaotic behaviour plays an important role given that
these systems are described by non-linear differential equations.
This chaotic behaviour gives rise to a rich spectrum of possible
evolutions and the surprising feature of predictable randomness
given that chaotic branches are deterministic (see for instance,
Refs. \cite{Coveney,Gleik}).

\vspace{0.3cm}

\noindent 5) Fom the discovery of the CP-symmetry violation in the
$K^0 - \bar{K}^0$ system, one infers from the CPT-theorem, a
fundamental cornerstone of quantum field theory, that the
T-symmetry is also violated. This means that on a quite elementary
level there exists an intrinsic irreversibility. The violation of
the CP-symmetry and also of baryon number in an expanding universe
are conditions from which the observed baryon asymmetry of the
universe (BAU) can be set (see for instance \cite{Buchmuller} and
references therein). An alternative route to explain the BAU is
through the violation of the CPT-symmetry itself \cite{Bertolami97}. 
This is possible, for instance, in the context of string theory \cite{Kost07}.

\vspace{0.3cm}

\noindent 6) The psychological perception of time is obviously irreversible and
historical. The past is recognizable and can be scrutinised, 
while the future is open and unknown.
This perception is presumably intimately related with the issue of
causation and the branching of possible outcomes towards the future. 
A fascinating related question is whether this irreversible psychological 
perception of time is the only one compatible with the laws of thermodynamics 
or whether it is the result of an advantageous evolutionary adaptation 
of our brain. The recent observation of a common cortical metrics of time, 
space and quantity \cite{Walsh03} might lead one to conjecture 
that the anatomical structure of our brain makes the associations of 
time with change and space with time rather natural. 
%It has been argued that the underlying
%philosophical basis for this can be found in Schopenhauer's theory
%of cognition \cite{Belenkiy08}.

\vspace{0.3cm}

\noindent 7) Systems bound gravitationally exhibit the so-called
gravito-thermal catastrophic behaviour \cite{Lynden-Bell}, meaning
that their entropy grows as they contract, which in turn implies
that their specific heat is negative. On the largest known scales,
the expansion of the Universe, which is itself adiabatic, is a
unique phenomenon, and as such, is conjectured to be the arrow of
time to which all others might be subordinated.

\section{Open issues}\label{Sec:V}

Let us briefly discuss here some problems related to the nature of
spacetime that remain essentially unsolved. These include a
putative correlation between the above discussed arrows of time
and the question of nonexistence of an explicit time variable in
the canonical Hamiltonian formulation of quantum gravity.

\subsection{Are the arrows of time correlated?}

One could argue that the existence of systems, from which a time
direction can be inferred, is not at all so surprising, as this
feature is the essence of all dissipative phenomena. It is possible
that this directional flow reflects, for instance, a particular
choice of boundary conditions which constrain the state of the
universe, rather than any restriction on its dynamics and
evolution. However, this point of view cannot account for the
remarkable fact that all known arrows of time do point from past to present, 
to present to future. In what follows, we briefly discuss some of the views
put forward to relate the arrows of time among themselves.
Thorough discussions can be found in Refs.
\cite{Davies,Zeh,GellMann}.

For instance, the philosopher Hans Reichenbach (1891 - 1953)
\cite{Reichenbach} argued in his book {\it The Direction of Time},
that the arrow of time in all macroscopic phenomena has its origin
in causality, which in turn should be the origin of the growth of
entropy. The argument is somewhat circular, but the suggested
connection is the most important point of the discussion. Possibly,
one the most original ideas about a putative correlation of time
arrows is due to the cosmologist Thomas Gold (1920 - 2004), who in
1958, suggested that all arrows of time should be subordinated to
the expansion of the Universe \cite{Gold}. This speculation gave
origin to some attempts, not quite successful, to correlate the
propagation of electromagnetic radiation to the expansion of the
Universe \cite{Hogarth,Hoyle}. Indeed, it is somewhat puzzling
that the found solutions indicate that retarded radiation is
compatible only with a steady-state universe, while advanced
radiation is found to be compatible only with evolutionary
universes (expanding or contracting ones). Clearly, these
solutions show evidence that the problem requires more complex and
realistic modelling.

A different starting of point is considered by Roger Penrose,
based on the Thermodynamics of Black Holes. It is suggested  that
the gravitational field must have an associated entropy which
should be measured by an invariant combination involving the Weyl
tensor \cite{Penrose1}. This idea allows for a consistent set up
for cosmology of the Generalized Second Principle of
Thermodynamics, as it arises in physics of black holes. This formulation
states that the Second Principle should apply to the sum of the
entropy of matter with the one of the black hole
\cite{Penrose2,HB}. The main point of Penrose's proposal is that
it resolves the paradox of an universe whose initial state is a
singularity or a black hole protected by a horizon, and hence with
an initial entropy that exceeds by many decades of magnitude the
entropy of the observed universe. Indeed, Penrose's suggestion
explains the low entropy of the initial state from its isotropy
and homogeneity as in this situation the Weyl tensor vanishes
\cite{Bertolami85}. The gravitational entropy will then increase
as the Weyl tensor grows as the universe becomes lumpier.

The growth of the total entropy can, in principle, account for the
asymmetry of psychological time as in this way the branching of
states and outcomes will occur towards the future.

Let us conclude this discussion with some comments on some recent
developments in the context of superstring/M-theory. These suggest
a multiverse approach of the {\it landscape} of vacua of the theory. This
corresponds to a {\it googleplexus} of $10^{500}$ vacua
\cite{BoussoPol}, which should be regarded as distinct universes.
Naturally, a suitable selection criteria for the vacuum that 
corresponds to our universe
should be found. If not, how then our universe has been chosen and emerged 
from the multitude of vacua of the theory?
Anthropic arguments \cite{Susskind} and quantum
cosmological considerations \cite{HMersini} have been proposed for
this purpose. For sure, these meta-theories of initial conditions
are not consensual, but they contribute to a better understanding
of the problem. Naturally, one should keep in mind that
non-perturbative aspects of string theory are poorly known
\cite{Polchinski}. The multiverse approach opens the possibility
of interaction among different universes \cite{Bertolami07}. This
interaction is suggested to be controlled by a Curvature Principle
and it is shown, in the context of a simple model of two
interacting universes, that the cosmological constant of one of
the universes can be driven to a vanishingly small value. The main
point of the argument is an action principle for the interaction
of universes using the curvature invariant $I_{i} = R_{\mu \nu
\lambda \sigma}^{i}  R^{\mu \nu \lambda \sigma}_{i}$, where $
R_{\mu \nu \lambda \sigma}^{i}$ is the Riemann tensor of each
universe. The proposed Curvature Principle also allows for a
possible solution for the entropy paradox of the initial state of
the universe \cite{Bertolami07}. From the point of view of another
universe, from which our universe can be perceived as if all its
mass were concentrated in some point and therefore, $I=48 M^2
r^{-6}$, where $r$ is the universe horizon's radius and $M$ its
mass - using units where $G=\hbar=c=1$. Thus, if the entropy
scales with the volume, then $S \sim r^3 \sim I^{-1/2}$; for the
case that the entropy scales according to the holographic
principle, suitable for AdS spaces \cite{Fischler,Bousso}, then $S
\sim r^2 \sim I^{-1/3}$. For both cases, given that $I \sim
\Lambda^2$ for the ground state, one obtains that $S \rightarrow
0$ in the early universe and, $S \rightarrow \infty$ when $\Lambda
\rightarrow 0$. The latter corresponds to the universe at late
time, which is consistent with the Generalized Second Principle of
Thermodynamics.

\subsection{Time in quantum gravity}

Quantum gravity is the theory that is expected to describe the
behaviour of space-time at distances of the order of the Planck
length, $L_P \simeq 10^{-35}~m$. It is still largely unknown, even
though important developments have been made in the context of the
superstring/M-theory, the most studied quantum gravity approach.
This approach leads to a quite rich set of ideas and concepts, but
has not provided satisfactory answers to some fundamental problems
such as for instance to account for the smallness of the
cosmological constant \cite{Witten00}. Furthermore, it exhibits
the vacuum selection problem discussed above, which seriously
compromise the predictability power of the whole programme.

In order to understand the conceptual difficulties that afflict
quantum gravity, let us discuss how the procedure of quantization
of gravity seriously challenge the well tested methods of quantum
field theory. Indeed, even though the metric, $g_{\mu
\nu}(\vec{r}, t)$, can be seen as a bosonic spin-two field, when
attempting to consider its quantization through an equal-time
commutation relation for the corresponding operator:
\begin{equation}
[\hat{g}_{\mu \nu} (\vec{r},t), \hat{g}_{\mu \nu}(\vec{r'},t)] = 0~,
\end{equation}
for $\vec{r}-\vec{r'}$ space-like, one faces a problem of
definition: i) To start with, in order to establish that
$\vec{r}-\vec{r'}$ is space-like, one must specify the metric; ii)
Being an operator relationship, it must hold for any state of the
metric; iii) Without a proper specification of the metric,
causality is ill-defined.

These obstacles suggest that one should consider instead the
canonical quantization procedure based on the Hamiltonian
formalism. In
this framework, one splits spacetime and selects {\it foliations}
where the physical degrees of freedom of the metric are the
space-like ones, $h_{ab}=^{(3)}g_{ab}$. The resulting Hamiltonian
is a sum of constraints, one associated with invariance under time
reparametrization, the others related with invariance under
3-dimensional diffeomorphisms. If one considers only Lorentzian
geometries (a quite restrictive condition !), then only the first
constraint is relevant. The solution of the classical constraint
is given by \cite{DeWitt}: 
\beq H_0=0~, \eeq where \beq H_0=\sqrt{h}\left[h^{-1}
\Pi_{ab} \Pi^{ab} - {1 \over 2} h^{-1} \Pi_{a} \Pi^{a} -
^{(3)}R\right]~, 
\eeq 
$h$ being the determinant of the 3-metric
$h_{ab}$, $\Pi_{ab} = {\delta L/\delta\dot{h_{ab}}}$, the
respective canonical conjugate momentum, obtained from the
Lagrangian of the problem and $^{(3)}R$ the 3-curvature. The
quantization consists in turning the momenta into operators for
some operator ordering and applying the resulting Hamiltonian
operator into a wave function, the wave function of the universe,
$\Psi[h_{ab}]$ \cite{DeWitt}: 
\beq \hat{H}_0 \Psi[h_{ab}] = 0~. 
\label{WDW} 
\eeq
This is the well known Wheeler-DeWitt equation, the starting point
of the so-called quantum cosmology
(see e.g. \cite{BMourao91} and references therein),
where the canonical approach
has been throughly used to study the initial conditions for the
Universe.

In the context of the canonical formalism, the problem of time
(see Ref. \cite{Isham} for a thorough discussion) arises as one
does not have a Schr\"odinger-type equation for the evolution of
states, but rather, the constrained problem (\ref{WDW}), where
time is one of the variables within $H_0$. Of course, this does
not mean that there is no evolution, but rather that there is no
straightforward way to obtain from the formalism a variable that
resembles the ``cosmic time'' that is employed in classical
cosmology. Furthermore, it is somewhat hasty to conclude in this
context that time is not a fundamental physical variable.

An attempt to solve this difficulty assumes a semi-classical
approach \cite{Vilenkin}, where time is identified with the
scale factor or some function of it, once the behaviour of the
metric is classical and the wave function of the universe admits a
WKB approximation. In this context, the Wheeler-DeWitt equation
can be written, at least in the minisuperspace approximation where
one admits only a finite (or an infinite but numerable) set of
degrees of freedom, as the Hamilton-Jacobi equation for the action
in the WKB approximation. Physically it implies that time can be
identified as such only after the metric starts behaving as a
classical variable.

Another interesting proposal is the so-called {\it Heraclitean time
proposal} \cite{UW,Bertolami95,Carroll08}. This is based on a suggestion
due to Einstein \cite{Einstein}, according to which the
determinant of the metric might not be a dynamical quantity. In
this approach, usually referred to as {\it unimodular gravity}, the
cosmological constant arises as an integration constant and an 
time variable can be introduced as the classical
Hamiltonian constraint assumes the form \cite{UW}: \beq H =
\Lambda h^{1/2}~, \eeq and thus, for a given space-like
hypersurface $\Sigma$, one can obtain a  Schr\"odinger-like
equation: \beq i{\partial \Psi \over \partial t}= \int_{\Sigma}
d^3 x h^{-1/2} \hat{H}_0 \Psi = \hat{H} \Psi~. \eeq For sure, the
problem of time in quantum gravity remains still an open problem
and the presented proposals are examples of possible lines for
future research.

\section{Conclusions}\label{Sec:Conclusion}

In this review paper we have examined the nature of time and its
relationship with causation. Particular attention has been paid to
the new feature of the special and the general theory of
relativity according to which time flows at different rates for
different observers. This is sharply contrasting with the
situation in Newtonian physics where time flows at a constant rate
for all observers, providing a notion of absolute time. It is
shown, in the context of special relativity, how the concept of
universal simultaneity is unattainable, and consequently, that the idea
of an universal present is impossible.

In this context, the Block Universe description emerges, a
formulation where all times, past and future are equally present,
and the notion of the flow of time is a subjective illusion. This
leads one to the possibility that time is indeed a dimension, and
not a process. Despite of the popularity of the Block Universe
representation in physics, most particularly in the relativistic
community, this viewpoint is still met with some suspicion. Notice
however, that for many scientists, Einstein included,
irreversibility is essentially an illusion.

A related issue about the nature of time concerns its flow, from
past to present, from present to future. The question here is the
mismatch between macrophysics, described by the Second Law of
Thermodynamics, and through which a distinct arrow of time arises
from the growth of entropy in irreversible processes, and
microphysics, whose fundamental evolution equations, classical and
quantum, are symmetric under time reversal. In the context of
statistical mechanics, it is assumed that the deterministic
behaviour in the microphysics context is lost as the macroscopic
description necessarily averages out the micro-properties of the
systems. As discussed above, another possibility is to assume that
the thermodynamic arrow of time is a consequence of the initial
conditions of the Universe which is inexorably correlated to the
direction of the Universe's expansion.

Finally, in what concerns the fundamental ontological nature of
time, we believe that it remains still an open question whether
time is a real fundamental quantity or, instead, a composite or a
convenient parameter to describe the physical laws and most
particularly to pose, in unequivocal terms, causation. Causation
that, as we have seen, is a most crucial feature not only in
physics, but according to some philosophers,
like David Hume (1711 - 1776),
also for the very human understanding of reality.

%-----------------------------------------------

%-----------------------------------------------
\end{document}